\documentstyle[epsfig]{mn}

\def\Mesz{M\'esz\'aros~}

\def\beq{\begin{equation}}
\def\enq{\end{equation}}
\def\bea{\begin{eqnarray}}
\def\ena{\end{eqnarray}}
\def\bec{\begin{center}}
\def\enc{\end{center}}
\def\etal{{et al.~}}
\def\eps{\epsilon}
\def\ic{{\rm IC}}

\def\ergcm2si{\hbox{ergs~cm$^{-2}$s$^{-1}$}}

\title[X-ray afterglows of gamma-ray bursts]{X-ray afterglows of gamma-ray bursts in the synchrotron
self-Compton dominated regime}
\author[Z. Li and L. M. Song]{Zhuo Li\thanks{E-mail: lizhuo@mail.ihep.ac.cn} and L. M. Song\\
Particle Astrophysics Lab., Institute of High Energy Physics,
Chinese Academy of Sciences, Beijing 100039, China}

\pubyear{2003}

\begin{document}

\maketitle

\begin{abstract}
We consider in this paper the effect of synchrotron self-Compton
process on X-ray afterglows of gamma-ray bursts. We find that for
a wide range of parameter values, especially for the standard
values which imply the energy in the electrons behind the
afterglow shock is tens times as that in the magnetic field, the
electron cooling is dominated by Compton cooling rather than
synchrotron one. This leads to a different evolution of cooling
frequency in the synchrotron emission component, and hence a
different (flatter) light curve slope in the X-ray range. This
effect should be taken into account when estimating the afterglow
parameters by X-ray observational data. For somewhat higher
ambient density, the synchrotron self-Compton emission may be
directly detected in X-ray range, showing varying spectral slopes
and a quite steep light curve slope.
\end{abstract}


\section{Introduction}
The current model (see reviews of Cheng \& Lu 2001 and \Mesz2002)
of gamma-ray burst (GRB) afterglows is that a sideways expanding
jet (Rhoads 1999) drives a blast wave propagating into the
circum-burst medium, and the shock-accelerated electrons give rise
to the afterglow emission. The main radiation process is believe
to be synchrotron emission by electrons (e.g., \Mesz \& Rees 1997;
Sari, Piran \& Narayan 1998), which is consistent with the
afterglow spectra (e.g., Galama \etal 1998). The polarization
detections in afterglows have also implicated the synchrotron
mechanism (e.g., Covino \etal 1999; Wijers \etal 1999). The
synchrotron self-Compton (SSC) emission is also important if the
energy density of radiated synchrotron photons exceeds that of the
magnetic field in the shock. This may be always the case since the
ratio between the post-shock energies in electrons and in the
magnetic field is commonly larger than $\sim10$, and a significant
fraction of the shock-heated electron energy is radiated away. The
SSC emission has been studied by previous works based on the
spherical afterglow model (Panaitescu \& \Mesz 1998; Wei \& Lu
1998; Totani 1998; Chiang \& Dermer 1999; Dermer, Chiang \& Mitman
2000; Dermer, B\"ottcher \& Chiang 2000; Panaitescu \& Kumar 2000;
Sari \& Esin 2001; Zhang \& M\'esz\'aros 2001). Two cases of X-ray
excess in the afterglow spectra have been explained to be the
inverse-Compton components (Harrison \etal 2001; Yost \etal 2002).

The observed X-ray emission from GRB afterglows usually comes from
synchrotron by fast cooling electrons, those electrons with
energy-lose times less than the dynamical time of the system. If
these electrons lose energy mainly by SSC rather than synchrotron,
more shock-heated electrons would cool rapidly, thus the
distribution of electrons and hence the X-ray light curve index
would be different from the synchrotron-dominated case
(Panaitescue \& Kumar 2001; Li, Dai \& Lu 2002). Furthermore, the
SSC component is possible to be directly detected in the X-rays in
some cases. Thus the SSC effects should be taken into account when
modelling the afterglow observational data. Panaitescu \& Kumar
(2002) had incorporated numerically the SSC mechanism in the
modelling of many GRB afterglows, in order to give out the
physical condition of relativistic jets in GRB afterglows.
However, most people still tend to use the simple asymptotic
relation of light curve, rather than the numerical modelling, to
fit the observational data. In many cases, though the afterglows
are in the SSC-dominated regime, the asymptotic relation in
synchrotron-dominated regime are still used in the fitting.
Therefore it is necessary to derive the parameter range in which
the afterglow are SSC-dominated, and then the analytical
asymptotic relation of afterglow light curve in this regime.

In this paper, we make more detailed study on the inverse-Compton
processes in GRB afterglows, especially the X-ray afterglow
emission from jets in SSC-dominated regimes. We first introduce in
section 2 the whole dynamical evolution of a beaming afterglow. In
section 3 we calculate the light curve of synchrotron emission and
the model parameter constraint on the SSC-dominated case. We then
discuss in section 4 the case when SSC emission emerges directly
in X-rays. Section 5 is a brief summary and discussion.

\section{Dynamical evolution}
Consider a beaming outflow from the GRB source, so called a jet,
which decelerates as sweeping up the ambient medium and sideways
expanding in the local sound speed. If the radiation energy is
negligible compared to the jet kinetic energy, the jet can be
regarded as adiabatic when considering its dynamic evolution. This
is always the case provided the energy fraction that goes into
shocked elections is $\eps_e\la0.1$, which is the common value
from model fit to observational data. For a higher $\eps_e$, the
jet will undergo first an early radiative stage, in which the
afterglow light curve index is relevant to $\eps_e$ (B\"ottcher \&
Dermer 2000; Li, Dai \& Lu 2002). We consider only adiabatic
dynamics here.

A jet, with equivalent isotropic energy $E$, coasts first with
initial Lorentz factor $\gamma_0$ until it sweeps up enough
material at a deceleration time
$t_0=(3E/32\pi\gamma_0^8nm_pc^5)^{1/3}$, with $n$ the ambient
medium density. After $t_0$ the jet begins to decelerate. The
deceleration of the jet includes three stages: First, when the
sideways expansion is not significant compared to the initial jet
open angle $\theta_0$, the jet undergoes a spherical-like phase
where the jet Lorentz factor decreases as $\gamma\propto t^{-3/8}$
(Blandford \& McKee 1976), with $t$ the observer's time, and the
jet open angle is $\theta\simeq\theta_0$. Secondly, when the
sideways expansion begins to dominate the dynamical evolution at
$t_j=t_0(\gamma_0\theta_0)^{8/3}$, we have $\theta\simeq
1/\gamma$, and the jet turns into a spreading phase where
$\gamma\propto t^{-1/2}$ (Rhoads 1999). Here we have assumed that
the sound speed in the relativistic stage is comparable to light
speed, $c_s\sim c$. Finally, the jet becomes non-relativistic,
$\gamma\approx1$, at $t_n=t_0^{3/4}t_j^{1/4}\gamma_0^2$ and
$\theta\sim 1$. In the non-relativistic phase the sideways
expansion is not important to affect the dynamical evolution, and
the shock's velocity $v\propto t^{-3/5}$, its radius $r\propto
t^{2/5}$.

The three special times are calculated in the following: \beq
t_0=90(E_{52}/n)^{1/3}\gamma_{0,2}^{-8/3}\mbox{s}, \enq \beq
t_j=4.2\times10^4(E_{52}/n)^{1/3}\theta_{0,-1}^{8/3}\mbox{s},
\enq\label{eq:tj} \beq
t_n=4.2\times10^6(E_{52}/n)^{1/3}\theta_{0,-1}^{2/3}\mbox{s},
\enq\label{eq:tn} where we have used the convention $U=10^xU_x$
and c.g.s units. Hereafter, by ``sphere" we mean the $t_0<t<t_j$
phase, by ``jet" the $t_j<t<t_n$ phase and by ``NR" the $t>t_n$
phase.


\section{Synchrotron emission}

The shock accelerates the ambient electrons to high energies, with
electron Lorentz factors described by a power-law distribution:
$dN_e/d\gamma_e\propto\gamma_e^{-p}$ for $\gamma_e>\gamma_m$. The
typical Lorentz factor of electrons is proportional to the
internal energy density of the shock as $\gamma_m\propto\gamma-1$.
At the beginning it is approximated as
$\gamma_m\approx610\eps_e\gamma_0$ at $t_0$, and evolves as
$\gamma_m \propto\gamma$ in the relativistic regime since
$\gamma-1\approx\gamma$, while in the NR phase it becomes
$\gamma_m\propto v^2\propto t^{-6/5}$ since $\gamma-1\propto v^2$
for NR. The magnetic field is also created by the shock, commonly
assumed to carry a fraction $\eps_B$ of the total internal energy
behind the shock front. Thus the energy density of magnetic field,
$B^2/4\pi$, is also proportional to $\gamma-1$. At the
deceleration time $t_0$, the magnetic field is
$B=(32\pi\eps_Bnm_pc^2)^{1/2}\gamma_0$, later on it evolves as
$B\propto\gamma$ in the relativistic regime [$\propto t^{-3/8}$
(sphere) and $\propto t^{-1/2}$ (jet)], while $B\propto v\propto
t^{-3/5}$ in NR phase.

Under these conditions the synchrotron radiation is produced, with
the instantaneous spectrum described as power-law segments (Sari,
Piran \& Narayan 1998). The typical frequency of synchrotron
photons is relevant to the typical electron energy,
\begin{equation}
\nu_m=\frac{x_pe}{\pi
m_ec}B\gamma_m^2\gamma\propto\left\{\begin{array}{ll}
    t^{-3/2}    &  {\rm sphere,}\\
    t^{-2}     &   {\rm jet,}\\
    t^{-3}     &   {\rm NR,}\end{array}\right.
\end{equation}
where $x_p$ is defined by Wijers \& Galama (1999) and of order of
unity.

The electrons lose energy through both synchrotron and SSC, and
the Compton parameter $Y$, i.e., the ratio between the
inverse-Compton to synchrotron luminosity, is calculated as (Sari
\& Esin 2001)
\begin{equation}\label{eq:Y}
Y=\frac{-1+\sqrt{1+4\eta\epsilon_e/\epsilon_B}}2\simeq \left\{
\begin{array}{ll}
  \eta\epsilon_e/\epsilon_B,  &{\rm if }~\eta\epsilon_e/\epsilon_B\ll 1,\\
  \sqrt{\eta\epsilon_e/\epsilon_B}, &{\rm if }~\eta\epsilon_e/\epsilon_B\gg 1,
                             \end{array} \right.
\end{equation}
where $\eta$ is the fraction of electron energy that is radiated
away (by both synchrotron and SSC). The synchrotron cooling
frequency, i.e. the frequency of the synchrotron photons radiated
by those electrons which cool on the dynamical time of the shock,
is given by
\begin{equation}
\nu_c=\frac{36\pi em_ec}{\sigma_T^2B^3\gamma t^2(1+Y)^2}.
\end{equation}
Since the electrons responsible to synchrotron frequencies above
$\nu_c$ lose energy quickly, the radiated fraction of electron
energy is therefore
\begin{equation}\label{eq:eta}
\eta=\left\{\begin{array}{ll}
  1 &   \hbox{for fast cooling, $\nu_c<\nu_m$,}\\
  (\nu_c/\nu_m)^{(2-p)/2}   &   \hbox{for slow cooling, $\nu_c>\nu_m$.}
                             \end{array} \right.
\end{equation}
The equations (\ref{eq:Y})-(\ref{eq:eta}) show that $Y$ and
$\nu_c$ are correlated, and these three equations should be
combined to solve the time evolutions of both $Y$ and $\nu_c$,
especially for the IC-dominated case, $Y>1$, which we focus on in
this paper. The $Y$ and $\nu_c$ should be solved by
numerical calculation, 
while for extreme case $Y\gg1$ we can reach an analytical result
(see also Li, Dai \& Lu 2002),
\begin{equation}
\nu_c\propto\left\{\begin{array}{ll}
    t^{-3/2+2/(4-p)}    &  {\rm sphere,}\\
    t^{-2+4/(4-p)}     &   {\rm jet,}\\
    t^{-3+28/[5(4-p)]}     &   {\rm NR.}\end{array}\right.
\end{equation}

The flux peaks at the lower one of the two frequencies $\nu_m$ and
$\nu_c$. The swept-up electron number is approximated by
$N_e\simeq\pi\theta^2r^3n/3$, and the power per unit time per unit
frequency emitted by single electron is (in the comoving frame)
$P_{\nu}=(3^{1/2}\phi_pe^3/m_ec^2)B$, where $\phi_p$ is calculated
by Wijers \& Galama (1999) and is of order of unity. Furthermore,
the energy emitted by total electrons is distributed over an area
of $\Delta S\sim \pi\theta^2D^2$ at a luminosity distance $D$ from
the source, the observed peak flux density is therefore
\begin{equation}
F_{\nu,\max}\simeq\frac{N_e\gamma P_{\nu}}{\Delta S}\propto
r^3\gamma B \propto\left\{\begin{array}{ll}
    {\rm const.}    &  {\rm sphere,}\\
    t^{-1}     &   {\rm jet,}\\
    t^{3/5}     &   {\rm NR.}\end{array}\right.
\end{equation}

Except for the very early times (see equation \ref{eq:tcm}), the
afterglow is generally in slow cooling regime with
$\nu_c\gg\nu_m$. We focus on the highest radiation energy range of
afterglows, i.e. the X-ray band, which usually corresponds to the
$\nu>\nu_c$ flux,
\begin{eqnarray}
F_{\nu>\nu_c}=F_{\nu,\max}(\nu_c/\nu_m)^{-(p-1)/2}(\nu/\nu_c)^{-p/2}\nonumber\\
\propto\left\{\begin{array}{ll}
    t^{-3p/4+1/(4-p)}    &  {\rm sphere,}\\
    t^{-p+(p-2)/(4-p)}     &   {\rm jet,}\\
    t^{-(66p-15p^2-52)/[10(4-p)]}     &   {\rm
    NR.}\end{array}\right.
\end{eqnarray}
This above equation expresses the light curve of synchrotron
emission in the IC-dominated case ($Y>1$)\footnote{Equation (28)
in Li, Dai \& Lu (2002) had shown the light curve scaling for the
frequency range of $\nu>\nu_c$ and in the jet spreading phase, but
that equation has a mistake in the $Y>1$ case.}. We summarize the
results together with previous works for synchrotron-dominated
case ($Y<1$) in table \ref{table}. Since the synchrotron emission
in the $\nu_m<\nu<\nu_c$ range,
$F_{\nu_m<\nu<\nu_c}=F_{\nu,\max}(\nu/\nu_m)^{-(p-1)/2}$, is
irrelevant to the evolution of  $\nu_c$, the light curve index in
this frequency range is the same as the synchrotron-dominated
case. The scaling relations for synchrotron-dominated case have
not been included in table \ref{table} and can be found in Sari,
Piran \& Halpern (1999) and Dai \& Lu (1999, 2000).

\subsection{Parameter range for strong Compton cooling} With different values of
physical parameters, e.g., $\eps_e$ and $\eps_B$, the system may
correspond to different cases of whether synchrotron- or
IC-dominated, therefore we discuss the parameter range now. In
general, the afterglow is initially in the fast cooling regime,
with $\nu_c<\nu_m$ and $\eta=1$, and then the Compton parameter is
a constant, $Y_0\approx\sqrt{\eps_e/\eps_B}$, provided commonly
$\eps_e>\eps_B$. It is not until a time,
\begin{equation}\label{eq:tcm}
t_{cm}=1.0\times10^3E_{52}n\eps_{e,-1}^2\eps_{B,-2}^2\left(1+\sqrt{\frac{\eps_{e,-1}}{\eps_{B,-2}}}\right)^2\mbox{s},
\end{equation}
that the afterglow becomes slow cooling and the Compton parameter
decreases as $Y\propto t^{-(p-2)/[2(4-p)]}$. For the cooling of
electrons to be still dominated by SSC process, the Compton
parameter at the point of jet break should be larger than unity:
$Y(t_j)>1$. This, with help of equation (\ref{eq:tj}), leads to
\begin{equation}
\eps_{e,-1}>0.24\frac{\theta_{0,-1}^{2/9}}{E_{52}^{1/18}n^{1/9}}\eps_{B,-2}^{2/3}
\end{equation}
for $p=2.2$ and
\begin{equation}
\eps_{e,-1}>0.45\frac{\theta_{0,-1}^{8/21}}{E_{52}^{2/21}n^{4/21}}\eps_{B,-2}^{3/7}
\end{equation}
for $p=2.4$. Thus, with the commonly taken parameters, such as
$\eps_e\sim0.1$ and $\eps_B\sim0.01$, the afterglow is still
Compton-dominated when the jet break in the light curve appears.

\begin{figure}
\centerline{\hbox{\psfig{figure=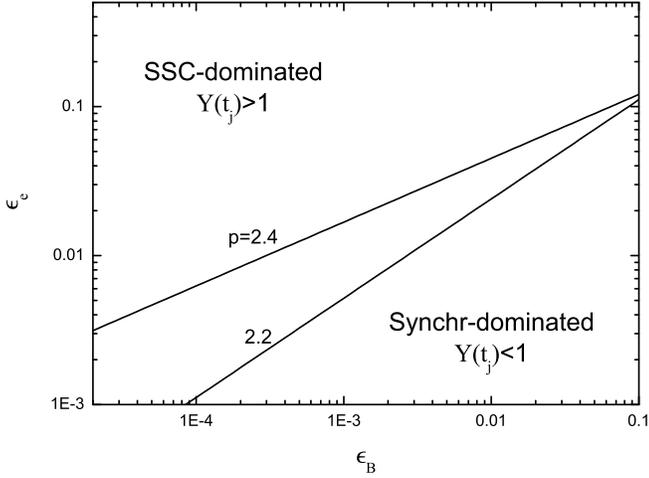,width=4in,angle=0}}}
\caption {Regions in the $\eps_e$, $\eps_B$ parameter space in
which synchrotron or SSC dominates when the jet break occurs. The
critical cases of $Y(t_j)=1$ are shown for $p=2.2$ and $p=2.4$.
The upper-left region is still SSC-dominated after jet break at
$t_j$, while the bottom-right region has become
synchrotron-dominated before $t_j$.}\label{fig:range}
\end{figure}

After the jet break point, the Compton parameter turns to drop
faster as $Y\propto t^{-(p-2)/(4-p)}$. If we require that the $Y$
value is still larger than unity when the jet goes into NR phase,
i.e., $Y(t_n)>1$, the condition is
\begin{equation}
\eps_{e,-1}>0.52\frac{\eps_{B,-2}^{2/3}}{E_{52}^{1/18}n^{1/9}\theta_{0,-1}^{1/9}}
\end{equation}
for $p=2.2$ and
\begin{equation}
\eps_{e,-1}>1.7\frac{\eps_{B,-2}^{3/7}}{E_{52}^{2/21}n^{4/21}\theta_{0,-1}^{4/21}}
\end{equation}
for $p=2.4$. Therefore the Compton cooling may dominate
synchrotron cooling even in the NR phase for the common parameter
values. So in the whole period of X-ray observation, Compton
cooling is strong. These above inequations are insensitive to the
initial condition of afterglows, like the total (isotropic) energy
$E$, the ambient density $n$ and the jet open angle $\theta_0$,
but sensitive to shock physics. We show the parameter ranges in
figure \ref{fig:range}.

\section{Direct detection of inverse-Compton component}
The SSC component dominates the synchrotron one in high enough
energy range, and its spectral shape can also be approximated by
broken power laws as synchrotron one (Panaitescu \& Kumar 2000;
Sari \& Esin 2001): $F_{\nu}^\ic\propto\nu^{1/3}$ for
$\nu<\min(\nu_m^\ic,\nu_c^\ic)$;
$F_{\nu}^\ic\propto\nu^{-(p-1)/2}$ for $\nu_m^\ic<\nu<\nu_c^\ic$
(or $F_{\nu}^\ic\propto\nu^{-1/2}$ for $\nu_c^\ic<\nu<\nu_m^\ic$);
and $F_{\nu}^\ic\propto\nu^{-p/2}$ for
$\nu>\max(\nu_m^\ic,\nu_c^\ic)$, where
$\nu_m^\ic\approx2\gamma_m^2\nu_m$ and
$\nu_c^\ic\approx2\gamma_c^2\nu_c$, with $\gamma_i$ being the
electron Lorentz factor corresponding to synchrotron frequency
$\nu_i$.

\begin{figure}
\centerline{\hbox{\psfig{figure=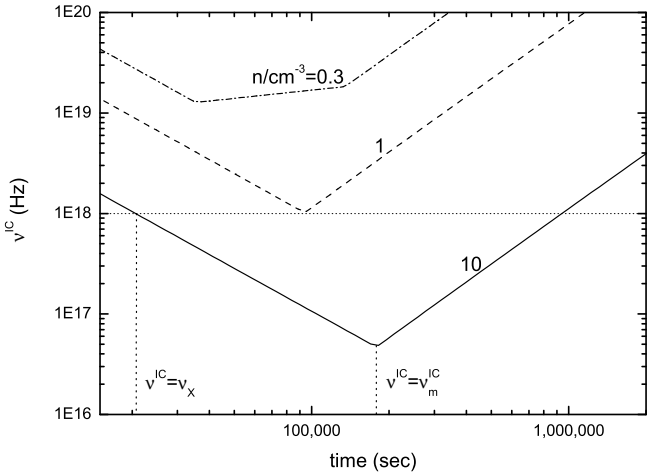,width=4in,angle=0}}}
\caption {The frequency above which the emission is dominated by
SSC, as function of time, for $n=0.3$, 1 and 10 cm$^{-3}$, using
$E=10^{53}$~ergs, $\theta_0=0.1$, $\gamma_0=150$, $p=2.4$,
$\eps_e=0.1$ and $\eps_B=10^{-3}$. The horizon line shows a X-ray
frequency $\nu_{\rm X}=10^{18}$~Hz. Only for the cases of
$n>1$~cm{-1} can $\nu^\ic$ drops below the X-ray band. For the
case of $n=10$~cm$^{-3}$ the special times of $\nu^\ic=\nu_{\rm
X}$ and $\nu^\ic=\nu_m^\ic$ are marked. }\label{fig:nuic}
\end{figure}

After a time $t_{cm}$ the system becomes slow cooling, with
$\nu_m<\nu_c$ for synchrotron component and $\nu_m^\ic<\nu_c^\ic$
for SSC component. If taken $\eps_e\sim0.1$ and $\eps_B\sim0.01$
typically, the system is in the SSC-dominated regime. Therefore we
here limit our discussion to the SSC-dominated ($Y>1$) and slow
cooling ($t_j>t_{cm}$) case, during which for typical parameters
the crossing point between the synchrotron and the SSC spectral
components, $\nu^\ic$, generally lies above the synchrotron
cooling frequency $\nu_c$ and below SSC cooling frequency
$\nu_c^\ic$. For the SSC emission to be detected directly in
X-rays, we need $\nu^\ic\la10^{18}$~Hz. This condition places a
lower limit on the ambient density (Sari \& Esin 2001). We
numerically calculate the emission by both synchrotron and SSC and
then the evolution of the crossing frequency $\nu^\ic$ with time
for different ambient densities, as show in figure \ref{fig:nuic}.
In general, the lower limit is $n>1$~cm$^{-3}$.

\begin{figure}
\centerline{\hbox{\psfig{figure=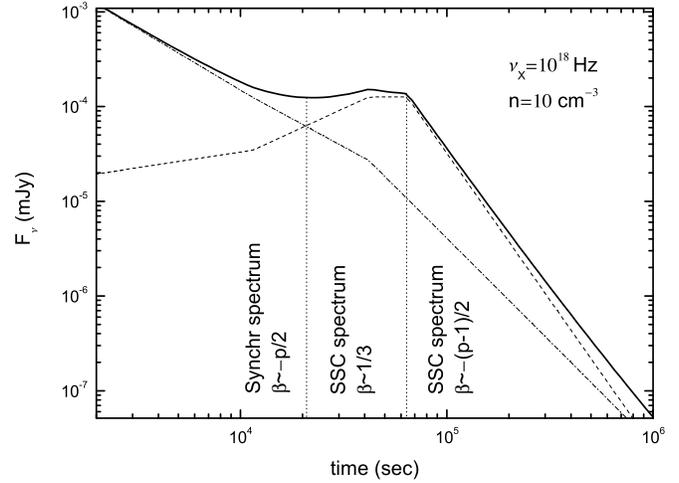,width=4in,angle=0}}}
\caption {X-ray ($\nu=10^{18}$~Hz) light curve in the case of
$n=10$~cm$^{-3}$. The other parameters are taken as:
$E=10^{53}$~ergs, $\theta_0=0.1$, $\gamma_0=150$, $p=2.4$,
$\eps_e=0.1$, $\eps_B=10^{-3}$ and $D=10^{28}$~cm. The total flux
({\it thick solid}) consists of synchrotron ({\it dashed}) and SSC
({\it dashed-dot}) components. At early times the X-ray flux is
dominated by synchrotron with spectral slope of $\nu^{-p/2}$.
Later, when $\nu^\ic$ moves into the X-ray band and the emission
is dominated by SSC, the flux rises/flattens and has a spectral
form of $\nu^{1/3}$ and then, when the $\nu_m^\ic$ drops into the
X-ray band, the flux decays fast and has a spectral form of
$\nu^{-(p-1)/2}$.}\label{fig:lc}
\end{figure}

In general, the $\nu_m^\ic$ moves into the X-ray band in the jet
spreading phase ($t_j<t<t_n$). For fixed X-ray frequency $\nu_{\rm
X}=10^{18}\nu_{18}$~Hz, the crossing time is $t_m^\ic\sim
5\times10^4\eps_{e,-1}^{4/3}\eps_{B,-2}^{1/6}
E_{52}^{1/3}\theta_{0,-1}^{2/3}n^{-1/6}\nu_{18}^{-1/3}$~s. Note
that we have assumed the slow cooling case which requires
$t_m^\ic>t_j>t_{cm}$. Around $t_m^\ic$ the observed flux evolves
as
\begin{equation}
F_{\nu_{\rm X}}^\ic\propto\left\{\begin{array}{ll}
    t^0\nu_{\rm X}^{1/3}    &  t_j<t<t_m^\ic,\\
    t^{-(3p-1)/2}\nu_{\rm X}^{-(p-1)/2}     &   t_m^\ic<t<t_n.\end{array}\right.
\end{equation}
The spectral slope changes gradually from 1/3 to $-(p-1)/2$, which
is different from the $-p/2$ slope in the high energy tail of
synchrotron component, and in the same time the light curve index
changes from zero to a steep decline. The relation between the
steep light curve index $\alpha_\ic$ ($F_{\nu}^\ic\propto
t^{-\alpha_\ic}$) and the spectral index $\beta_\ic$
($F_{\nu}^\ic\propto \nu^{-\beta_\ic}$) in the rapid decline is:
\begin{equation}
\alpha_\ic-3\beta_\ic-1=0.
\end{equation}
We emphasize that a steep light curve together with a shallow
spectral slope in X-ray band (e.g. $F_\nu^\ic\propto
t^{-3.1}\nu^{-0.7}$ for $p=2.4$ ) may imply the direct detection
of SSC-dominated emission component. Figure \ref{fig:lc} has shown
a case when SSC dominates the X-ray emission in jetted afterglows.

\section{Summary and discussion}
We have discussed in this paper the effect of SSC process on the
X-ray afterglow. For a wide range of parameter values (see figure
\ref{fig:range}), including the commonly taken ones
$\eps_e\simeq0.1$ and $\eps_B\simeq0.01$, the electron cooling is
dominated by IC cooling rather than synchrotron one. This leads to
a different evolution of cooling frequency $\nu_c$ in the
synchrotron emission component, and hence a different (flatter)
synchrotron light curve slope above $\nu_c$, say, the X-ray range.
The light curve index of jet-spreading phase in SSC-dominated
($Y>1$) case is flatter by a factor of $(p-2)/(4-p)$ than
synchrotron-dominated case. This SSC effect should be taken into
account when modelling in detail the X-ray observational data. It
should be noticed that in many case we should use the
SSC-dominated $\alpha-\beta$ relations (in table \ref{table})
rather than the synchrotron-dominated ones to fit the observation.

For somewhat higher ambient density, $n\ga3$~cm$^{-3}$, the SSC
emission dominates the synchrotron in X-ray range and can be
detected directly (see also Sari \& Esin 2001). The SSC light
curve shows a slope of $\alpha_\ic=2.5-3.4$ for $p=2-2.6$, quite
steeper than the synchrotron one. When the SSC component emerges,
the X-ray spectral slope varies, which may be detected by
observation.

The upcoming Swift satellite is due to launch at the end of 2003,
which is expected to catch more than 200 afterglows per year.
Owing to its rapid response, many afterglows may be rapidly
observed in O/UV and X-rays within one minute. The current
operating X-ray satellites, Chandra and XMM-Newton, have high
sensitive and spectral resolution. So many more detailed X-ray
observations of GRB afterglows are expected. We emphasize that the
X-ray observation of afterglows may help to follow the cooling of
electrons and help to investigate the SSC characteristics of
afterglows.

\section*{Acknowledgments}

This work was supported by the Special Funds for Major State Basic
Research Projects and by the National Natural Science Foundation
of China.

\newpage
\begin{table*}
\caption{The synchrotron light-curve index $\alpha$
($F_{\nu}\propto t^{-\alpha}$) as function of $p$ in the range of
$\nu>\nu_c$. The parameter-free relation between $\alpha$ and the
spectral index $\beta$ ($F_\nu\propto\nu^{-\beta}$) is given for
each case by substituting $p=2\beta$ as for $\nu>\nu_c$. The
numerical factors in the bracket correspond to $p=2.4$.}
\begin{tabular}{|c||c|c|c|}
\hline & \multicolumn{3}{|c|}{light curve index $\alpha$
($F_{\nu}\propto t^{-\alpha}$)}\\
& sphere & jet & non-relativistic \\
\hline\hline
& $\alpha=3(p-1)/4$ & $\alpha=p$ & $\alpha=(3p-4)/2$ \\
{$\nu>\nu_c$, $Y<1$} & $\alpha=3\beta/2-1/2$
& $\alpha=2\beta$ & $\alpha=(6\beta-4)/2$ \\
& (1.05) & (2.4) & (1.6) \\ \hline
& $\alpha=3p/4-1/(4-p)$ & $\alpha=p-(p-2)/(4-p)$ & $\alpha=(66p-15p^2-52)/[10(4-p)]$ \\
{$\nu>\nu_c$, $Y>1$} & $\alpha=3\beta/2-1/(4-2\beta)$ &
$\alpha=2\beta-(\beta-1)/(2-\beta)$ &
$\alpha=(66\beta-30\beta^2-26)/[10(2-\beta)]$ \\
& (1.18) & (2.15) & (1.25) \\
\hline
\end{tabular}
\label{table}
\end{table*}



\begin{thebibliography}{99}
\bibitem{} Blandford R. D., McKee, C. F., 1976, Phys. Fluids, 19, 1130
\bibitem{} B\"ottcher M., Dermer C. D., 2000, ApJ, 532, 281
\bibitem{} Cheng K. S., Lu T., 2001, Chin. J. Astron. Astrophys., 1, 1
\bibitem{} Chiang J., Dermer C. D., 1999, ApJ, 512, 699.
\bibitem{} Covino S. et al., 1999, A\&A, 348, L1
\bibitem{} Dai Z. G., Lu T., 1999, ApJ, 519, L155
\bibitem{} Dai Z. G., Lu T., 2000, ApJ, 537, 803
\bibitem{} Dermer C. D., B\"ottcher M., Chiang J., 2000, ApJ, 537, 255
\bibitem{} Dermer C. D., Chiang J., Mitman K. E., 2000, ApJ, 537, 785
\bibitem{} Galama T. J., Wijers R. A. M. J., Bremer M., Groot P.
J., Strom R. G., Kouveliotou C., van Paradijs J., 1998, ApJ, 500,
L97
\bibitem{} Harrison F. A. et al. 2001, ApJ, 559, 123
\bibitem{} Li Z., Dai Z. G., Lu T., 2002, MNRAS, 330, 955
\bibitem{} \Mesz P., 2002, ARA\&A,  40, 137
\bibitem{} \Mesz P., Rees M. J., 1997, ApJ, 476, 232
\bibitem{} Panaitescu A., {M\'esz\'aros} P., 1998, ApJ, 501, 772
\bibitem{} Panaitescu A., Kumar P., 2000, ApJ, 543, 66
\bibitem{} Panaitescu A., Kumar P., 2001, ApJ, 554, 667
\bibitem{} Panaitescu A., Kumar P., 2002, ApJ, 571, 779
\bibitem{} Rhoads J., 1999, ApJ, 525, 737
\bibitem{} Sari R., Esin A. A., 2001, ApJ, 548, 787
\bibitem{} Sari R., Piran T., Halpern J., 1999, ApJ, 524, L43
\bibitem{} Sari R., Piran T., Narayan R., 1998, ApJ, 497, L17
\bibitem{} Totani T., 1998, ApJ, 502, L13
\bibitem{} Wei D. M., Lu T., 1998, ApJ, 505, 252
\bibitem{} Wijers R. A. M. J. et al., 1999, ApJ, 523, L33
\bibitem{} Wijers R. A. M. J., Galama T. J., 1999, ApJ, 523, 177
\bibitem{} Yost S. A. et al., 2002, ApJ, 577, 155
\bibitem{} Zhang B., M\'esz\'aros P., 2001, ApJ, 559, 110
\end{thebibliography}
\end{document}